# Analyzing Uncertainty Matrices Associated with Multiple *S*-parameter Measurements

Nick M. Ridler, *Fellow, IEEE*, and Martin J. Salter, *Member, IEEE*

National Physical Laboratory, Teddington, UK   nick.ridler@npl.co.uk

*Abstract* — **This paper presents a detailed analysis of uncertainty matrices (i.e. measurement covariance matrices) associated with multiple complex-valued microwave scattering (*S*-) parameter measurements. The analysis is based on forming combinations of (2 × 2) sub-matrices from selected elements in the uncertainty matrix and using these sub-matrices to define uncertainty ellipses in two-dimensional measurement planes. These uncertainty ellipses can be used to assess the quality, accuracy and validity of the measurements. A simple example is given using measurements made on a waveguide mismatched line section at frequencies from 75 GHz to 110 GHz. The analysis is also useful when multiple *S*-parameter measurements are used as input quantities for measurement models used to determine other measurement quantities (i.e. as output quantities).**

## I. INTRODUCTION

Since the publication, in 1993, of the ISO "Guide to the Expression of Uncertainty in Measurement" (GUM) [1], much work has been undertaken, in many areas of science and technology, to implement the general techniques outlined in the GUM.[1] In the area of high-frequency electromagnetic metrology, a key issue with implementing the techniques given in the GUM has been that many of the measurands that are encountered are complex-valued quantities, i.e. the measurand has both a Magnitude and Phase, or, equivalently, a Real and Imaginary component.

A procedure for dealing with the uncertainty in a complex-valued measurand has been described in [3, 4], where the uncertainties in the Real and Imaginary components of the measurand were analysed in terms of a (2 × 2) covariance matrix. The examples given in [3, 4] were of *S*-parameter measurements that occur in microwave metrology. The matrix formulation that was used enabled the correlation between the uncertainties in the Real and Imaginary components of the complex-valued measurand to be included in the overall expression of uncertainty. This led to the use of the correlation coefficient to describe the interrelationship between the two components of the measurand.

More recently, the procedures outlined in the GUM [1, 2] have been extended to deal with measurement situations where there are multiple output quantities (i.e. where there is more than one measurand) [5]. This extended procedure enables the interactions between all the measurands to be taken into account when establishing the uncertainty in the multiple output quantities. Correlation coefficients can be established between suitable pairs of output quantities selected from the complete set of output quantities. This leads to a correlation matrix which has the same number of elements as the uncertainty matrix. The elements in the correlation matrix are the correlation coefficients that describe all possible correlations between pairs of output quantities.

This paper applies the techniques given in [5] to the situation where there are multiple complex-valued output quantities. The techniques given in [3, 4] that accounted for the interactions between the Real and Imaginary components of an individual complex-valued *S*-parameter are extended to take account of interactions between the components of all the complex-valued *S*-parameters. This includes interactions between components of different *S*-parameters – i.e. the interaction between a component of one *S*-parameter and a component of a different *S*-parameter. For example, the interaction between the real (or imaginary) component of $S_{11}$ and the real (or imaginary) component of, say, $S_{21}$, and so on.

Accounting for interactions between different *S*-parameters can be important especially in situations when the behavior of one *S*-parameter is physically related to the behavior of other *S*-parameters. For example, a measurement device that contains a significant mismatch will generate both an increase in reflection and a decrease in transmission, compared to the matched condition. The relationship between the amount of energy reflected and the amount of energy transmitted is governed by the law of conservation of energy – i.e. energy that is reflected cannot at the same time be transmitted, so, in this situation, there is clearly a strong physical relationship between reflection and transmission. For an *n*-port device (where $n > 1$) containing significant mismatch, there is expected to be correlation between the reflection coefficients, $S_{ii}$, $(i = 1, \ldots, n)$ and the transmission coefficients, $S_{ij} (i = 1, \ldots, n; j = 1, \ldots, n; i \neq j)$.

---

[1] Since 1993, there have been several minor revisions/updates to the GUM. The most recent edition was published in 2008 [2]. This edition can be downloaded from www.bipm.org/en/publications/guides.

In general, for measurements of multi-port devices, there may be many possible combinations of pairs of components of different $S$-parameters. However, in this paper, the investigation is restricted to measurements of two-port devices, as this is sufficient to illustrate the necessary concepts.[2] A simple example is given showing measurements made on a mismatched waveguide line in the WR-10 waveguide size, developed by OML, Inc. As mentioned above, the mismatch in this device should give rise to significant interactions between the reflection and transmission coefficients. This is investigated in terms of the interactions between the various combinations of the components of these complex-valued coefficients.

## II. CONCEPT

The $S$-parameters for a two-port device can be represented using the following scattering matrix:

$$[S] \equiv \begin{bmatrix} S_{11} & S_{21} \\ S_{12} & S_{22} \end{bmatrix} \tag{1}$$

The uncertainty for each of the four complex-valued $S$-parameters in (1) can be represented by a $(2 \times 2)$ uncertainty matrix[3]. For example, the uncertainty matrix for $S_{11}$ is:

$$\begin{pmatrix} u^2(S_{11_R}) & u(S_{11_R}, S_{11_I}) \\ u(S_{11_I}, S_{11_R}) & u^2(S_{11_I}) \end{pmatrix} \tag{2}$$

where the sub-subscripts R and I are used to indicate the Real and Imaginary components, respectively, of an $S$-parameter, $S_{ij}$ ($i = 1, 2; j = 1, 2$). This matrix represents the uncertainties in the Real and Imaginary components of $S_{11}$, including the interaction between these components, using the off-diagonal element $u(S_{11_R}, S_{11_I})$, or, equivalently $u(S_{11_I}, S_{11_R})$. Similar $(2 \times 2)$ matrices are used to represent the uncertainties in, and interactions between, the Real and Imaginary components of $S_{21}$, $S_{12}$ and $S_{22}$.

In addition, other uncertainty terms are needed to represent interactions between components of different $S$-parameters. This results in an $(8 \times 8)$ uncertainty matrix being used to represent all the uncertainty information for the four $S$-parameter measurements of a two-port device, where off-diagonal elements are used to represent all the interactions between the components of these four $S$-parameters. The elements of this $(8 \times 8)$ uncertainty matrix are listed in Table 1.

The shaded regions in Table 1 indicate the four $(2 \times 2)$ sub-matrices that represent the uncertainty associated with each of the four $S$-parameters, independent of any interactions between other $S$-parameters.

Without loss of generality, this investigation is further restricted to consider measurement of two-port devices that exhibit both reciprocity (i.e. $S_{21} = S_{12}$) and symmetry ($S_{11} = S_{22}$), as this will simplify the analysis but will still be sufficient to illustrate the necessary concepts.[4] Under these conditions, only the $S$-parameters, $S_{11}$ and $S_{21}$, need to be considered. The $(8 \times 8)$ uncertainty matrix described in Table 1 can therefore be reduced to include only terms involving uncertainty in $S_{11}$ (Real and Imaginary), or $S_{21}$ (Real and Imaginary), or a combination of both $S_{11}$ and $S_{21}$ (Real and Imaginary). The $(8 \times 8)$ uncertainty matrix therefore reduces to the following $(4 \times 4)$ uncertainty matrix:

$$\begin{pmatrix} u^2(S_{11_R}) & u(S_{11_R}, S_{11_I}) & u(S_{11_R}, S_{21_R}) & u(S_{11_R}, S_{21_I}) \\ u(S_{11_I}, S_{11_R}) & u^2(S_{11_I}) & u(S_{11_I}, S_{21_R}) & u(S_{11_I}, S_{21_I}) \\ u(S_{21_R}, S_{11_R}) & u(S_{21_R}, S_{11_I}) & u^2(S_{21_R}) & u(S_{21_R}, S_{21_I}) \\ u(S_{21_I}, S_{11_R}) & u(S_{21_I}, S_{11_I}) & u(S_{21_I}, S_{21_R}) & u^2(S_{21_I}) \end{pmatrix} \tag{3}$$

This reduction process results in six sub-matrices representing the interactions between the Real and Imaginary components of $S_{11}$ and $S_{21}$. Two of these sub-matrices are the usual uncertainty matrices for $S_{11}$ and $S_{21}$ − i.e. the $S_{11}$ sub-matrix:

$$\begin{pmatrix} u^2(S_{11_R}) & u(S_{11_R}, S_{11_I}) \\ u(S_{11_I}, S_{11_R}) & u^2(S_{11_I}) \end{pmatrix} \tag{4}$$

relates to the Real and Imaginary components of $S_{11}$ and describes the uncertainty associated with a measurement coordinate $(S_{11_R}, S_{11_I})$ in the $S_{11}$ two-dimensional complex plane.

Similarly, the $S_{21}$ sub-matrix:

$$\begin{pmatrix} u^2(S_{21_R}) & u(S_{21_R}, S_{21_I}) \\ u(S_{21_I}, S_{21_R}) & u^2(S_{21_I}) \end{pmatrix} \tag{5}$$

relates to the Real and Imaginary components of $S_{21}$ and describes the uncertainty associated with a measurement coordinate $(S_{21_R}, S_{21_I})$ in the $S_{21}$ two-dimensional complex plane.

However, the other four sub-matrices are uncertainty matrices that represent the uncertainties in, and interactions between, a component of one $S$-parameter and a component of a different $S$-parameter. For example, for the Real component of $S_{11}$ and the Real component of $S_{21}$, the sub-matrix:

---

Table 1: Elements of the (8 × 8) uncertainty matrix associated with the S-parameter matrix in (1)

| | | $S_{11}$ | | $S_{21}$ | | $S_{12}$ | | $S_{22}$ | |
|---|---|---|---|---|---|---|---|---|---|
| | | Real | Imag | Real | Imag | Real | Imag | Real | Imag |
| $S_{11}$ | Real | $u^2(S_{11_R})$ | $u(S_{11_R},S_{11_I})$ | $u(S_{11_R},S_{21_R})$ | $u(S_{11_R},S_{21_I})$ | $u(S_{11_R},S_{12_R})$ | $u(S_{11_R},S_{12_I})$ | $u(S_{11_R},S_{22_R})$ | $u(S_{11_R},S_{22_I})$ |
| | Imag | $u(S_{11_I},S_{11_R})$ | $u^2(S_{11_I})$ | $u(S_{11_I},S_{21_R})$ | $u(S_{11_I},S_{21_I})$ | $u(S_{11_I},S_{12_R})$ | $u(S_{11_I},S_{12_I})$ | $u(S_{11_I},S_{22_R})$ | $u(S_{11_I},S_{22_I})$ |
| $S_{21}$ | Real | $u(S_{21_R},S_{11_R})$ | $u(S_{21_R},S_{11_I})$ | $u^2(S_{21_R})$ | $u(S_{21_R},S_{21_I})$ | $u(S_{21_R},S_{12_R})$ | $u(S_{21_R},S_{12_I})$ | $u(S_{21_R},S_{22_R})$ | $u(S_{21_R},S_{22_I})$ |
| | Imag | $u(S_{21_I},S_{11_R})$ | $u(S_{21_I},S_{11_I})$ | $u(S_{21_I},S_{21_R})$ | $u^2(S_{21_I})$ | $u(S_{21_I},S_{12_R})$ | $u(S_{21_I},S_{12_I})$ | $u(S_{21_I},S_{22_R})$ | $u(S_{21_I},S_{22_I})$ |
| $S_{12}$ | Real | $u(S_{12_R},S_{11_R})$ | $u(S_{12_R},S_{11_I})$ | $u(S_{12_R},S_{21_R})$ | $u(S_{12_R},S_{21_I})$ | $u^2(S_{12_R})$ | $u(S_{12_R},S_{12_I})$ | $u(S_{12_R},S_{22_R})$ | $u(S_{12_R},S_{22_I})$ |
| | Imag | $u(S_{12_I},S_{11_R})$ | $u(S_{12_I},S_{11_I})$ | $u(S_{12_I},S_{21_R})$ | $u(S_{12_I},S_{21_I})$ | $u(S_{12_I},S_{12_R})$ | $u^2(S_{12_I})$ | $u(S_{12_I},S_{22_R})$ | $u(S_{12_I},S_{22_I})$ |
| $S_{22}$ | Real | $u(S_{22_R},S_{11_R})$ | $u(S_{22_R},S_{11_I})$ | $u(S_{22_R},S_{21_R})$ | $u(S_{22_R},S_{21_I})$ | $u(S_{22_R},S_{12_R})$ | $u(S_{22_R},S_{12_I})$ | $u^2(S_{22_R})$ | $u(S_{22_R},S_{22_I})$ |
| | Imag | $u(S_{22_I},S_{11_R})$ | $u(S_{22_I},S_{11_I})$ | $u(S_{22_I},S_{21_R})$ | $u(S_{22_I},S_{21_I})$ | $u(S_{22_I},S_{12_R})$ | $u(S_{22_I},S_{12_I})$ | $u(S_{22_I},S_{22_R})$ | $u^2(S_{22_I})$ |

$$\begin{pmatrix} u^2(S_{11_R}) & u(S_{11_R},S_{21_R}) \\ u(S_{21_R},S_{11_R}) & u^2(S_{21_R}) \end{pmatrix} \quad (6)$$

describes the uncertainty associated with a measurement coordinate $\left(S_{11_R},S_{21_R}\right)$ in a two-dimensional plane with the Real component of $S_{11}$ as one axis and the Real component of $S_{21}$ as the other axis. This is not a conventional $S$-parameter. Instead, it is a two-dimensional quantity based on a combination of two components coming from two different complex-valued $S$-parameters.

In the same way, for the Real component of $S_{11}$ and the Imaginary component of $S_{21}$, the sub-matrix:

$$\begin{pmatrix} u^2(S_{11_R}) & u(S_{11_R},S_{21_I}) \\ u(S_{21_I},S_{11_R}) & u^2(S_{21_I}) \end{pmatrix} \quad (7)$$

describes the uncertainty associated with a measurement coordinate $\left(S_{11_R},S_{21_I}\right)$ in a two-dimensional plane with the Real component of $S_{11}$ as one axis and the Imaginary component of $S_{21}$ as the other axis.

For the Imaginary component of $S_{11}$ and the Real component of $S_{21}$, the sub-matrix:

$$\begin{pmatrix} u^2(S_{11_I}) & u(S_{11_I},S_{21_R}) \\ u(S_{21_R},S_{11_I}) & u^2(S_{21_R}) \end{pmatrix} \quad (8)$$

describes the uncertainty associated with a measurement coordinate $\left(S_{11_I},S_{21_R}\right)$ in a two-dimensional plane with the Imaginary component of $S_{11}$ as one axis and the Real component of $S_{21}$ as the other axis.

Finally, for the Imaginary component of $S_{11}$ and the Imaginary component of $S_{21}$, the sub-matrix:

$$\begin{pmatrix} u^2(S_{11_I}) & u(S_{11_I},S_{21_I}) \\ u(S_{21_I},S_{11_I}) & u^2(S_{21_I}) \end{pmatrix} \quad (9)$$

describes the uncertainty associated with a measurement coordinate $\left(S_{11_I},S_{21_I}\right)$ in a two-dimensional plane with the Imaginary component of $S_{11}$ as one axis and the Imaginary component of $S_{21}$ as the other axis.

## III. EXAMPLE

To illustrate the above concepts, measurements were made on a mismatched waveguide component designed and developed by OML, Inc. The waveguide size for this device is WR-10 and so measurements were made at frequencies from 75 GHz to 110 GHz. These measurements are used to present the data shown in Figures 1 to 3.

Figure 1 shows the magnitudes of $S_{11}$ and $S_{21}$. These are the conventional $S$-parameters for such a device. The magnitudes of these $S$-parameters show the characteristic undulation in response due to interacting mismatches inside the device. Maximum reflection (and minimum transmission) occurs at approximately 78 GHz, 87 GHz, 95 GHz and 104 GHz. Minimum reflection (and maximum transmission) occurs at approximately 75 GHz, 83 GHz, 91 GHz, 100 GHz and 109 GHz. As mentioned previously, the strong relationship between the behavior of the reflection and the transmission is due to energy conservation – i.e. signal that is reflected cannot be transmitted, and vice versa.

The strong relationship, shown in Figure 1, between the value of $S_{11}$ and $S_{21}$ at any given frequency can be investigated further by examining the behavior of combinations of components of both $S_{11}$ and $S_{21}$ – these combinations are shown in Figures 2 and 3. In Figure 2, the magnitude of the vectors $(S_{11_R}, S_{21_R})$ and $(S_{11_I}, S_{21_I})$ are plotted as a function of frequency, and in Figure 3, the magnitude of the vectors $(S_{11_R}, S_{21_I})$ and $(S_{11_I}, S_{21_R})$ are plotted as a function of frequency. The choice of which vector to plot on which of these graphs is not important – however, the choice made here does indicate a strong relationship between combinations of the components of $S_{11}$ and $S_{21}$. (In principle, the magnitudes of all six voltage ratios could be plotted on the same graph, but having this amount of data displayed on the same graph might make it difficult to interpret.)

At each frequency, in Figures 1, 2 and 3, the uncertainty in the six vectors, $(S_{11_R}, S_{11_I})$, $(S_{21_R}, S_{21_I})$, $(S_{11_R}, S_{21_R})$, $(S_{11_R}, S_{21_I})$, $(S_{11_I}, S_{21_R})$ and $(S_{11_I}, S_{21_I})$, can be represented by the $(2 \times 2)$ uncertainty matrices shown in (4) to (9), respectively.

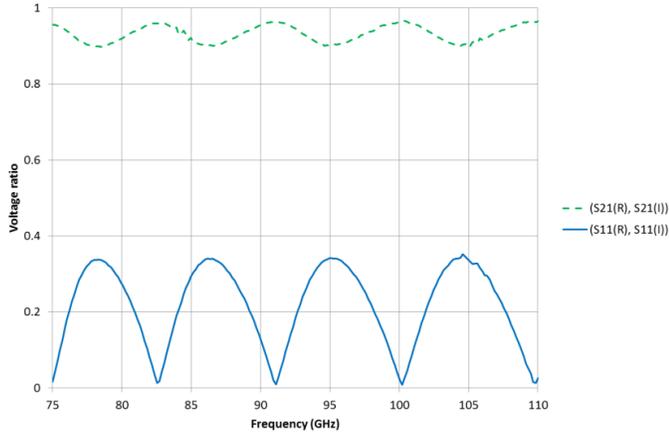

Figure 1: Magnitude of $S_{11}$ and $S_{21}$, derived from the $S$-parameter measurements

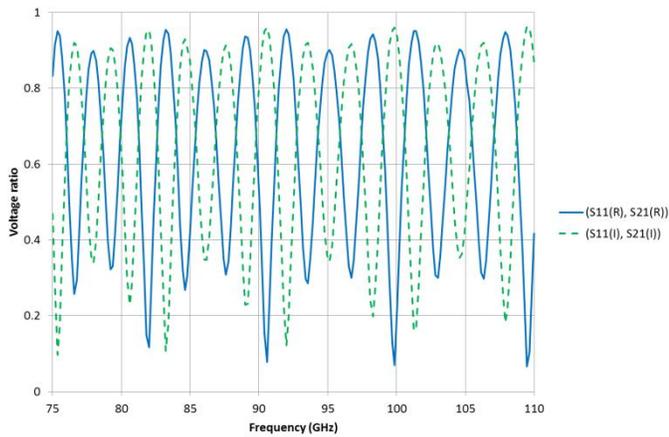

Figure 2: Magnitude of the vectors $(S_{11_R}, S_{21_R})$ and $(S_{11_I}, S_{21_I})$, derived from the $S$-parameter measurements

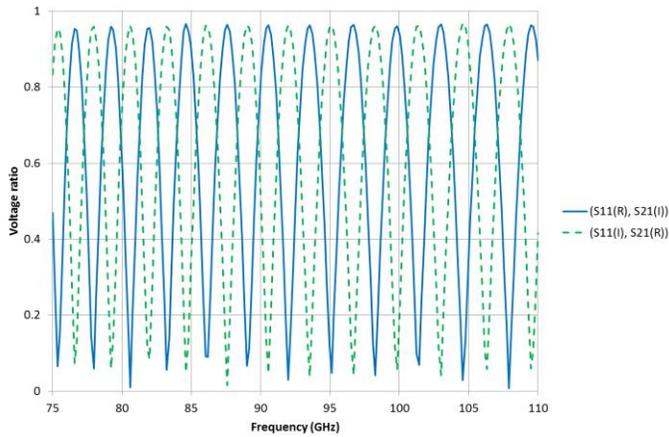

Figure 3: Magnitude of the vectors $(S_{11_R}, S_{21_I})$ and $(S_{11_I}, S_{21_R})$, derived from the $S$-parameter measurements

## IV. Uncertainty ellipses

If $S = (x, y)^T$ denotes a variable point in the measurement plane[5], $\bar{S} = (\bar{x}, \bar{y})^T$ denotes the measured value of the two-dimensional vector-valued measurand at a particular frequency and

$$V = \begin{pmatrix} s^2(\bar{x}) & s(\bar{x}, \bar{y}) \\ s(\bar{y}, \bar{x}) & s^2(\bar{y}) \end{pmatrix} \tag{10}$$

denotes the uncertainty matrix (covariance matrix) associated with the measured value, then the following equation represents an ellipse that bounds a region of uncertainty in the measurement plane associated with the measured value

$$(S - \bar{S})^T V^{-1} (S - \bar{S}) = k^2 \tag{11}$$

where $k$ is a "coverage factor" [2].[6]

It has been shown elsewhere [7] that the elements of a $(2 \times 2)$ covariance matrix can be used to construct an ellipse in a two-dimensional plane used to depict a two-dimensional vector measurand. For an $(2 \times 2)$ uncertainty matrix, this ellipse represents a region of uncertainty associated with the measured value of the two-dimensional measurand.

On this occasion, the measurement procedure for the WR-10 mismatch waveguide line included making a series of 12 replicate measurements of the device, in order to assess the repeatability of the measurements. These replicate measurements were used to establish the elements in the uncertainty matrices described in (4) to (9).[7] This procedure was applied at each measurement frequency. In addition, uncertainty ellipses were constructed using these uncertainty matrices, (4) to (9), for each of the six two-dimensional vectors described in the previous section. Sets of six ellipses were generated at each measurement frequency. As an example, the six ellipses that were produced at 75 GHz are shown in Figures 4 to 6. In all these Figures, a large coverage (i.e. multiplying) factor, $k$, has been used ($k = 500$) to enable the ellipses to be seen easily. The red dashed circle in each of the plots is the "unit circle" centred on the origin of the plane.

---

[5] The use of a superscript "T" here indicates the transpose of a vector.

[6] Note that in equation (11), superscript "-1" indicates the inverse of a matrix.

[7] These uncertainty matrices therefore do not contain information concerning all uncertainty contributions to these measurements. However, these matrices are considered sufficient to demonstrate the concepts described in this paper.

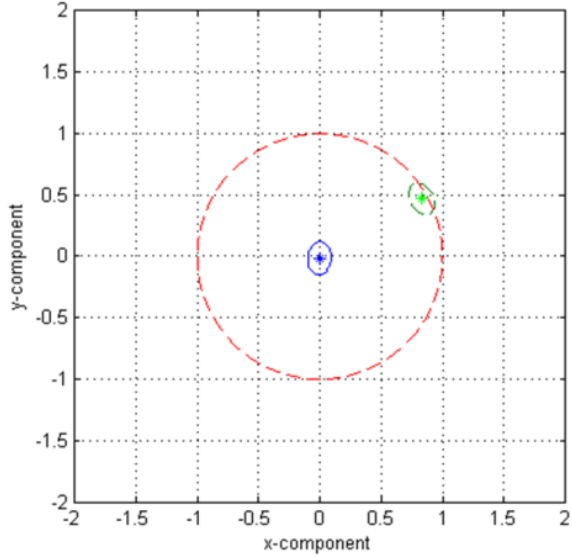

Figure 4: Uncertainty ellipses at 75 GHz. The solid blue ellipse is for the vector $(S_{11_R}, S_{11_I})$ and the dashed green ellipse is for the vector $(S_{21_R}, S_{21_I})$

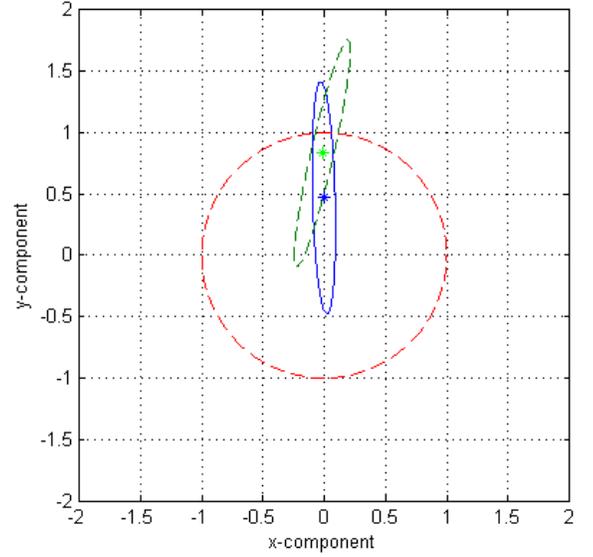

Figure 6: Uncertainty ellipses at 75 GHz. The solid blue ellipse is for the vector $(S_{11_R}, S_{21_I})$ and the dashed green ellipse is for the vector $(S_{11_I}, S_{21_R})$

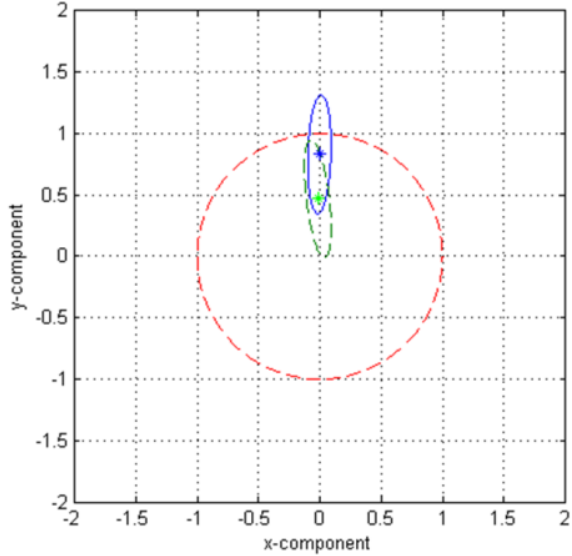

Figure 5: Uncertainty ellipses at 75 GHz. The solid blue ellipse is for the vector $(S_{11_R}, S_{21_R})$ and the dashed green ellipse is for the vector $(S_{11_I}, S_{21_I})$

## V. DISCUSSION

The uncertainty ellipses shown in Figures 4 to 6 provide a visual representation of the interactions between the components of the measured $S$-parameters. Figure 4 relates directly to $S_{11}$ and $S_{21}$, respectively, whereas Figures 5 and 6 show the interactions between a component of $S_{11}$ (either Real or Imaginary) and a component of $S_{21}$ (either Real or Imaginary).

These uncertainty ellipses can be used to assess the quality of the measurement results in terms of accuracy – i.e. relatively large ellipses suggest relatively poor measurement quality, and vice versa. The uncertainty ellipses can also be used to verify measurements by comparison with other sets of measurement data (see, for example, [8]). A detailed and rigorous verification procedure for multiple $S$-parameter measurement data is beyond the scope of this paper, and so this will form the subject of a subsequent paper.

## VI. CONCLUSION

A procedure has been presented for the detailed analysis of the uncertainty associated with measurements of multiple complex-valued $S$-parameters. The procedure follows recommendations, outlined in Supplement 2 to the GUM [5], for the treatment of uncertainty for any number of output quantities (i.e. measurands). The procedure quantifies the uncertainty for each individual $S$-parameter using a ($2 \times 2$) uncertainty matrix (i.e. a measurement covariance matrix). Additional ($2 \times 2$) matrices are then used to quantify the interactions between components of different $S$-parameters.

All these $(2 \times 2)$ uncertainty matrices are effectively sub-matrices drawn from a 'parent' uncertainty matrix which contains all uncertainty information for the multiple $S$-parameter measurements. For a two-port device, this parent matrix is a fully populated $(8 \times 8)$ uncertainty matrix – more generally, for an $n$-port device, the parent matrix will be a $(2n^2 \times 2n^2)$ uncertainty matrix.

For a reciprocal, symmetric, two-port device (as presented in this paper), six ellipses are needed, at each measurement frequency, to describe all the interactions between the uncertainty components for the two $S$-parameters, $S_{11}$ and $S_{21}$. For a non-reciprocal, non-symmetric, two-port device, a total of 28 ellipses would be needed to describe all interactions between the uncertainty components for all the four $S$-parameters, $S_{11}$, $S_{21}$, $S_{12}$ and $S_{22}$.

The analysis presented in this paper can also be applied to the situation where measurements of multiple $S$-parameters are used as inputs to the determination of other subsequent measurement quantities. The use of fully populated $(2n^2 \times 2n^2)$ uncertainty matrices for measurements of $n$-port devices enables the effects due to all interactions between the components of the $S$-parameters to be taken into account, in terms of their effects on these subsequent measurement quantities.

ACKNOWLEDGEMENT

The authors would like to thank Dr Yuenie Lau (OML, Inc) for the loan of the mismatched waveguide line that was used for the experimental results presented in this work. The work described in this paper was jointly funded by the European Metrology Research Programme (EMRP) Project SIB62 'HF-Circuits', and, the National Measurement System Directorate of the UK government's Department for Business, Innovation and Skills. The EMRP is jointly funded by the EMRP participating countries within EURAMET and the European Union.

REFERENCES

[1] "Guide to the Expression of Uncertainty in Measurement", International Organization for Standardization, 1st edition, ISBN 92-67-10188-9, 1993.

[2] JCGM 100:2008, "Evaluation of measurement data – Guide to the expression of uncertainty in measurement".

[3] N M Ridler and M J Salter, "Evaluating and expressing uncertainty in complex S-parameter measurements", *Proc. 56th ARFTG Microwave Measurement conference*, Boulder, CO, USA, November 2000.

[4] N M Ridler and M J Salter, "An approach to the treatment of uncertainty in complex *S*-parameter measurements", *Metrologia*, Vol 39, No 3, pp 295-302, June 2002.

[5] JCGM 102:2011, "Evaluation of measurement data – Supplement 2 to the 'Guide to the expression of uncertainty in measurement' – Extension to any number of output quantities".

[6] JCGM 101:2008, "Evaluation of measurement data – Supplement 1 to the 'Guide to the expression of uncertainty in measurement' – Propagation of distributions using a Monte Carlo method".

[7] N M Ridler and P R Young, "Investigating the random error distribution of vector network analyser measurements", *Proc. 29th ARMMS Conference Digest*, pp 26-34, Bracknell, UK, April 1999.

[8] K Kuhlmann and R Judaschke, "Reduction technique for Measurement Comparisons with Complex-valued Measurands", *Proc. Conference on Precision Electromagnetic Measurements (CPEM)*, Rio de Janeiro, Brazil, August 2014.